\documentclass[aps,prx,twocolumn,superscriptaddress,floatfix,longbibliography]{revtex4-1}
\usepackage{bm,amsmath,amsfonts,amssymb,braket}
\usepackage{latexsym,dsfont,array,layout,mathrsfs,textcase}
\usepackage{graphicx,graphics,subfigure,xcolor}
\usepackage{comment,verbatim}
\usepackage{float}
\usepackage{times}
\usepackage{xspace}
\usepackage{stmaryrd}
\usepackage{multirow}
\usepackage{mathtools}
\usepackage{booktabs}
\usepackage[colorlinks,linkcolor=blue,citecolor=blue,urlcolor=blue]{hyperref}
\usepackage{epstopdf}
\usepackage{ulem}

\renewcommand{\Re}{\operatorname{Re}}
\DeclareMathOperator\arctanh{arctanh}

\begin{document}
\title{Exact solution of the Bose-Hubbard model with unidirectional hopping}
\author{Mingchen Zheng}
\affiliation{Beijing National Laboratory for Condensed Matter Physics, Institute of
Physics, Chinese Academy of Sciences, Beijing 100190, China}
\affiliation{School of Physical Sciences, University of Chinese Academy of Sciences,
Beijing 100049, China}
\author{Yi Qiao}
\affiliation{Institute of Modern Physics, Northwest University, Xi'an 710127, China}
\author{Yupeng Wang}
\affiliation{Beijing National Laboratory for Condensed Matter Physics, Institute of
Physics, Chinese Academy of Sciences, Beijing 100190, China}
\author{Junpeng Cao}
\affiliation{Beijing National Laboratory for Condensed Matter Physics, Institute of
Physics, Chinese Academy of Sciences, Beijing 100190, China}
\affiliation{School of Physical Sciences, University of Chinese Academy of Sciences,
Beijing 100049, China}
\affiliation{Songshan lake Materials Laboratory, Dongguan, Guangdong 523808, China}
\affiliation{Peng Huanwu Center for Fundamental Theory, Xian 710127, China}
\author{Shu Chen}
\email{schen@iphy.ac.cn}
\affiliation{Beijing National Laboratory for Condensed Matter Physics, Institute of
Physics, Chinese Academy of Sciences, Beijing 100190, China}
\affiliation{School of Physical Sciences, University of Chinese Academy of Sciences,
Beijing 100049, China}

\date{\today}

\begin{abstract}
A one-dimensional Bose-Hubbard model with unidirectional hopping is shown to be exactly solvable. Applying the algebraic Bethe ansatz method, we prove the integrability of the model and derive the Bethe ansatz equations. The exact eigenvalue spectrum can be obtained by solving these equations.
The distribution of Bethe roots reveals the presence of a superfluid-Mott insulator transition at the ground state, and the critical point is determined. By adjusting the boundary parameter, we demonstrate the existence of a non-Hermitian skin effect even in the presence of interaction, but it is completely suppressed for the Mott insulator state in the thermodynamical limit. Our result represents a new class of exactly solvable non-Hermitian many-body systems, which has no Hermitian correspondence and can be used as a benchmark for various numerical techniques developed for non-Hermitian many-body systems.

\end{abstract}

\maketitle


\textit{Introduction.---}
Exact solutions of integrable quantum many-body models, e.g., the Lieb-Liniger model \cite{Lieb2,Lieb21}, Yang-Gaudin model \cite{Yang,Gaudin} and Fermi Hubbard model \cite{Lieb-Wu},
provide crucial insights into the understanding of correlation effect and have wide application in cold atomic and condensed matter physics \cite{GuanXW}.
However, constructing a new quantum integrable model with simple form and clear physical meanings is a difficult and fascinating mission.
Recently a great deal of progress has been made in the theory of non-Hermitian physics \cite{Bender98,Bender07,Bergholtz,Ashida}.
Some novel phenomena such as the enriched non-Hermitian topological classification \cite{Gong,Sato,Zhou,LiuCH1,LiuCH2,LiuCH3},
non-Hermitian skin effect (NHSE) \cite{13,15,16,17,18,LeeCH}, and scale-free localization \cite{Murakami,LiLH2021,GuoCX2023,WangZ2023,Bergholtz2023} have attracted intensive studies.
At present, most novel phenomena and concepts about the non-Hermitian effects are built based on noninteracting systems.
Then it is necessary to demonstrate whether these concepts are applicable and what new effect arises when the interaction is considered. Recently, there is attracting growing interest in exploring non-Hermitian phenomena in many-body systems, e.g., the interplay of non-Hermitian skin effect and interaction \cite{ZhangSB,ZhangDW,Alsallom}, the fate of the correlated phase in the presence of non-Hermiticity \cite{XuZ,PanL,ZhangDW,Nori2020,Fukui,HeL}, and non-Hermitian topological phases in correlated systems \cite{Hatsugai2022,Hatsugai2020,Ryu}.

Most of the existing studies on the non-Hermitian many-body systems are carried out by numerical diagonalization, which suffers from the many-body exponential wall problem and the problem of numerical errors and calculation precision in the diagonalization of non-Hermitian systems \cite{Colbrook,GuoCX2021}. The current understanding of NHSE and scale-free localization in correlated systems is quite limited. Exact solutions of non-Hermitian integrable many-body systems can provide benchmarks for understanding novel phenomena due to the interplay of non-Hermiticity and interaction.
In this work, we propose an integrable non-Hermitian Bose-Hubbard model describing the interacting bosons in a chain with unidirectional hopping,
which is one of conserved quantities constructed from the expansion coefficients of transfer matrix, whose integrability is guaranteed by the Yang-Baxter equation.
It is well known that the Bose-Hubbard model is not integrable and cannot be analytically solved \cite{Haldane,Fisher}. Compared to the previously reported exactly solvable non-Hermitian many-body systems \cite{Ueda2021,Fukui,PanL,Andrei,WangZ2301,PanL2022}, whose integrability is inherited from their Hermitian corresponding models by either introducing an imaginary gauge field or complex continuation of parameters, our model is a new integrable system without a Hermitian counterpart.
By using an algebraic Bethe ansatz method \cite{Korepin,WYCS}, we exactly solve the model and obtain the exact energy spectrum. 
Although the spectrum of a non-Hermitian many-body system is generally complex, we find that the ground state (defined by the minimum of real parts of eigenvalues) of our model is real. We unveil the occurrence of a superfluid-Mott insulator transition for integer filling cases by adjusting the ratio of interaction strength and hopping amplitude. Additionally, based on the obtained exact eigenstates, we also demonstrate the existence of non-Hermitian skin effect even in the presence of interaction, which is, however, found to be completely suppressed for the Mott state in the thermodynamical limit.

\textit{Model and its exact solution.---}
The model Hamiltonian reads
\begin{equation}\label{H1}
H=-t \left[ \sum_{j=1}^{N-1}b_j^{\dag}b_{j+1}+\epsilon b_N^{\dag}b_1 \right]+\frac{U}{2}\sum_{j=1}^N n_j(n_j-1),
\end{equation}
where ${b}_{j}^{\dagger }$ and ${b}_{j}$ are bosonic creation and annihilation operators at $j$th site respectively,
${n}_{j}= {b}_{j}^{\dag } {b}_{j}$ is the particle number operator on the $j$th site,
$N$ is the number of sites, $t$ quantifies the hopping between two sites with the nearest neighbor,
$\epsilon$ is the boundary parameter, and $U$ characterizes the strength of on-site interaction.
For simplicity, we consider $t>0$ and the on-site interaction is repulsive, i.e., $U>0$.
Since the Hamiltonian includes only a unidirectional hopping term, it is a non-Hermitian model.

First we prove that the model (\ref{H1}) is integrable in the framework of the quantum inverse scattering method \cite{Korepin}. A series of  conserved quantities of the Hamiltonian (\ref{H1})
can be generated by a transfer matrix
\begin{equation}
t(u)=\text{tr}_0[L_{0,N}(u) L_{0,N-1}(u) \cdots L_{0,1}(u) K_0],
\end{equation}
where $u$ is the spectral parameter, $\text{tr}_0$ means the trace in the two-dimensional auxiliary space $V_0$,
the subscripts $\{1,...,N\}$ denote the physical spaces, $L_{0,j}(u)$ is
the Lax operator defined in the tensor space $V_{0}\otimes V_j$,
and $K_0$ is the $2\times 2$ diagonal matrix given by $ K_0= \text{diag}(1, \epsilon)$.
In the auxiliary space, the Lax operator can be expressed by the matrix
\begin{equation}
L_{0,j}(u)=\begin{pmatrix}
u-n_{j} &g b_{j}\\g b_{j}^{\dag}& -g^2
\end{pmatrix}, \label{n-1}
\end{equation}
where the elements are the bosonic operators defined in the $j$th site and $g$ is a constant.
The Lax operator \eqref{n-1} satisfies the Yang-Baxter relation
\begin{equation}
    R_{0,\bar{0}} (u-v) {{L}}_{0,j} (u){{L}}_{\bar{0},j} (v)={{L}}_{\bar{0},j} (v){{L}}_{0,j} (u) R_{0,\bar{0}}(u-v),\label{n-2}
\end{equation}
where $R_{0,\bar{0}} (u)$ is the $R$ matrix defined in the auxiliary spaces 0 and $\bar{0}$ with the form of
\begin{equation}
    R_{0,\bar{0}} (u)=\begin{pmatrix}
    u-1& 0 & 0 & 0\\
    0 & u & -1 & 0\\
    0 & -1 & u & 0\\
    0 & 0 & 0 & u-1 \end{pmatrix}.
    \end{equation}

Using the Yang-Baxter relation \eqref{n-2}, it can be proven that the transfer matrices with different spectral parameters commute with each other, i.e., $[ t(u), t(v)]=0$.
Expanding the transfer matrix $t(u)$ with respect to $u$,
\begin{equation}
t(u)=u^{N}+C_{1}  u^{N-1}+C_{2} u^{N-2}+\cdots,
\end{equation}
then all the expansion coefficients $C_n$ $(n=1,...,N)$ are commutative and can be taken as conserved quantities.
Choosing one of them or a certain combination of them as a Hamiltonian $H$, then $[H, C_n] = 0$ and the model is integrable.
Direct calculation gives the operators $C_1$ and $C_2$ as
\begin{eqnarray}
&&C_{1}=-\sum_{j=1}^{N} n_{j}, \\
&&C_{2}= \frac {1}{2} \sum_{i\neq j}^{N} n_{i} n_{j}+g^2 \left(\sum_{j=1}^{N-1} b_{j}^{\dag}b_{j+1} + \epsilon b_{N}^{\dag}b_{1}\right).
\end{eqnarray}
The integrable Hamiltonian \eqref{H1} is constructed as
\begin{equation}
H=\frac{t}{2g^2}\left(C_1^2+C_1-2 C_2 \right),\label{n-3}
\end{equation}
where the on-site interaction is parameterized as $U=t/g^2$.

Using the algebraic Bethe ansatz and considering the number of bosons to be $M$, we obtain the eigenvalues of the transfer matrix $t(u)$ as \cite{SM}
\begin{equation}
\Lambda(u)=u^N \prod_{j=1}^{M}\frac{u-\beta_{j}/U-1}{u-\beta_{j}/U}+\epsilon (-t/U)^N \prod_{j=1}^{M}\frac{u-\beta_{j}/U+1}{u-\beta_{j}/U},
\end{equation}
where the $M$ Bethe roots $\{\beta_j\}$ should satisfy the Bethe ansatz equations (BAEs)
\begin{equation}\label{BAE2}
\Big(\frac{\beta_j}{-t}\Big)^N= \epsilon \prod_{l\neq j}^M \frac{\beta_j-\beta_l+U}{\beta_j-\beta_l-U}, \quad j=1,\cdots, M.
\end{equation}
Taking the logarithm of BAEs \eqref{BAE2}, we have
\begin{equation}\label{BAE4}
N \ln\Big(\frac{\beta_j}{-t\sqrt[N]{\epsilon}}\Big)=\sum_{l\neq j}^M \Theta \Big(\frac{U}{\beta_j-\beta_l}\Big)+2 i\pi I_j,
\end{equation}
where $\Theta (z)\equiv 2\arctanh (z)$ and $I_j$ is the quantum number.
Each set of integers $\{I_j|j=1, 2, ..., M\}$ characterizes one eigenstate.
The eigenvalue of Hamiltonian is
\begin{equation}
E=\sum_{j=1}^M \beta_j .\label{n-5}
\end{equation}
The solutions of BAEs \eqref{BAE2} or \eqref{BAE4} determine the energy spectrum of the Hamiltonian \eqref{H1} completely.
\begin{figure*}[tbp]
\includegraphics[width=1\textwidth]{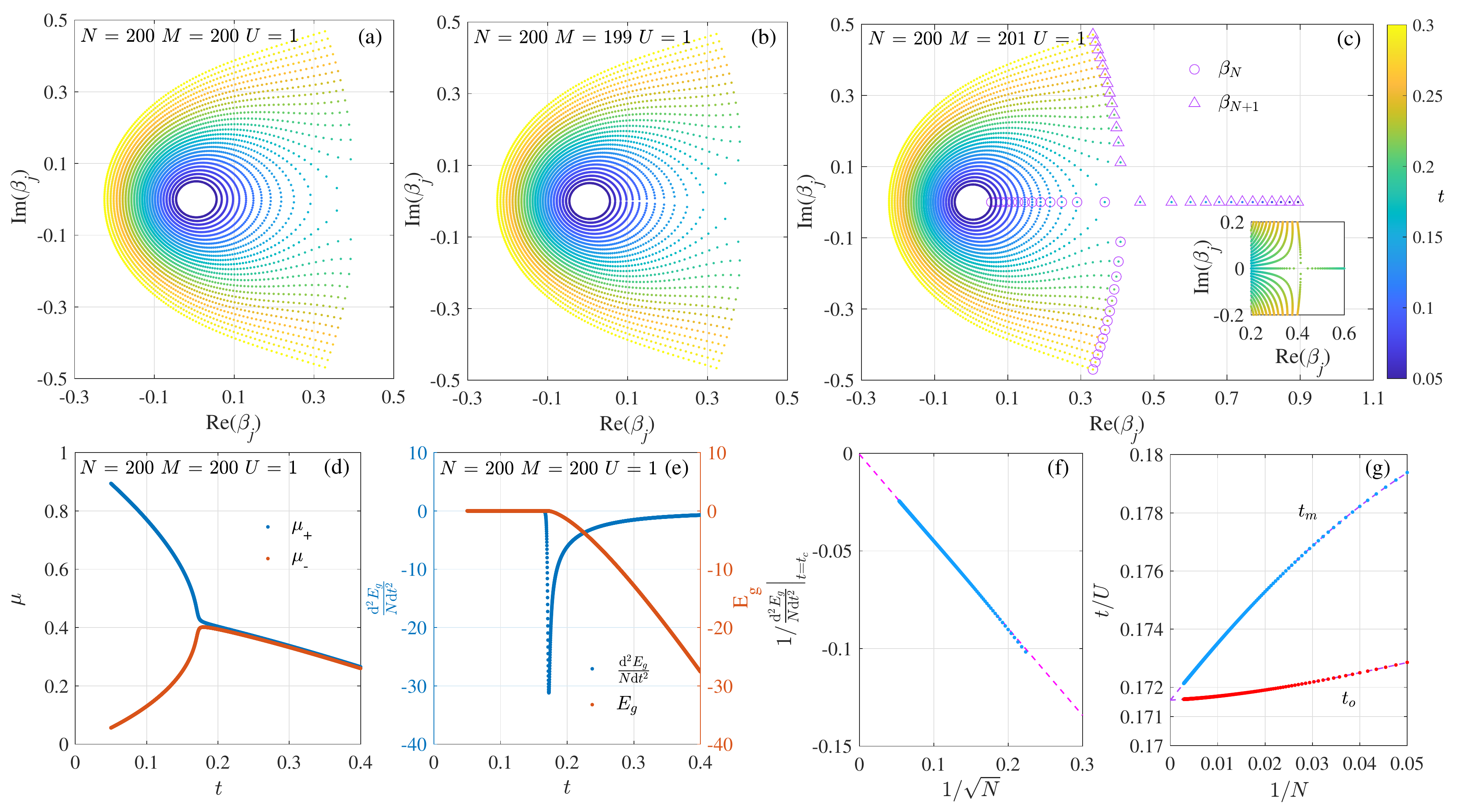}
\caption{(a)-(c) Distribution of Bethe roots of the ground state for $U=1,\ N=200$, and $M=\ 200\ $ (a),$\ 199\ $(b),$\ 201\ $(c)
with $t$ varying from 0.3 to 0.05 (the step size is 0.01). The inset in (c) displays a part of the Bethe roots with a finer step size of 0.001.
(d) Numerical results of $\mu_+$ and $\mu_-$ for the system of $N=M=200$ and $U=1$ with $t$ varying from 0.4 to 0.05 (step size 0.0001).
(e) The ground state energy $E_g$ (red point) and the second derivative of $E_g/N$ with respect to $t/U$ (blue point) for $N=200$ and $M=200$ with $t$ varying from 0.4 to 0.05 (step size 0.0001).
(f) The extreme values of ($d^2 E_g/N d t^2$) plotted against $1/\sqrt{N}$ from $N=20$ to $N=350$ ($\rho=1$).
The dashed fitted line is $1/(d^2 E_g/N d t^2)=C_1/\sqrt{N}+C_2$. The fitting parameters are $C_1=-0.44558$ and $C_2=-4.9649\times 10^{-4}$.
(g) The value of $t_m$ (blue points) and $t_o$ (red points) versus $1/N$ with $N=20-350$ ($\rho=1$).
The dashed lines are fitted to determine the values of $t_m$ and $t_o$ as the system size tends to infinity.
Fitting results with $N \rightarrow \infty$ give $t_m=0.17155$ and $t_o=0.17156$.}
\label{fig1}
\end{figure*}

{\it The Superfluid-Mott transition.--}
First, we consider the case of $\epsilon=1$.  At the limit of $U=0$, the Bethe roots take
$\beta_j=t\exp[i\pi(N+2 l)/N]$ with $l=1,...,N$.
When $U \neq 0$, the Bethe roots can be obtained by solving BAEs \eqref{BAE2}.
To verify the accuracy of our analytical results, we calculate the solutions of BAEs \eqref{BAE2} and, thus, obtain all the eigenvalues for a small system with $N=4$ and $M=4$.
We compare the results of BAEs with exact diagonalization  and confirm that they are consistent (see Supplemental Material \cite{SM}).

For the Bose-Hubbard model with integer filling, the system undergoes a superfluid-Mott insulator transition as the ratio of $U/t$ is varied \cite{Greiner}.
The critical value of $U/t$ has been determined using various methods such as mean field theory, quantum Monte Carlo simulation \cite{Scalettar91,Batrouni92}, and density matrix renormalization group method \cite{Kuhner98}.

Now we explore the superfluid-Mott phase transition for the Bose-Hubbard model with unidirectional hopping.
We shall focus on the case with unity filling factor $\rho=M/N=1$ and study its ground state properties.
At $U=0$, the ground state of the system is characterized by $N$ Bethe roots located at $\beta_j=-t$, indicating the condensation of $N$ bosons.
As $U/t$ increases from zero, the Bethe roots gradually move away from this point, forming a curve on the complex plane. When $U/t$ exceeds a critical value, the curve closes up and the Bethe roots fill a single ring. To visualize this behavior, we plot the distribution of the Bethe roots of the ground state for various $t$ in Fig.\ref{fig1} (a) while fixing $U=1$. It can be seen that as $t$ decreases, the arc of the curve gradually expands until it closes.
When $U/t$ is sufficiently large, the Bethe roots distribute uniformly on a circle with radius $t$, which is consistent with the hard-core limit of $U/t \rightarrow \infty$.

When a boson is removed from the system, we have $M=N-1$, and the distributions of the Bethe roots are similar to those in the case of $M=N$, as shown in Figure \ref{fig1} (b). However, adding a boson to the system, i.e., $M=N+1$, leads to a markedly different behavior in the distribution of Bethe roots when $U/t$ exceeds a critical value. As shown in Figure \ref{fig1} (c), the two roots with the largest real part, denoted as $\beta_N$ and $\beta_{N+1}$, approach the real axis as $t$ decreases.
At $t=t_o$, they coincide. For $t<t_o$, they repel each other and move oppositely along the real axis. If $U/t$ is large enough, we have $\beta_{N+1} \approx \beta_N + U - 2t$, and the other $N$ Bethe roots are distributed on a circle similar to the $M=N$ case.

Next we reveal that the change in the patterns of Bethe root across $t_o$ is a key indicator of the superfluid-Mott transition.
Define $\mu_+=E_N(\rho N+1)-E_N(\rho N)$ and $\mu_-=E_N(\rho N)-E_N(\rho N-1)$,
where $E_N(M)$ is the ground state energy for the system with $M$ bosons and $N$ sites.
If the system is in the Mott phase, there exists a nonzero gap $\Delta \mu=\mu_+-\mu_->0$, whereas the excitation in the superfluid phase is gapless.
As the parameter $U/t$ changes, the position where the energy gap opens or closes marks the superfluid-Mott transition point.
In the thermodynamic limit $N$ $(M)\to \infty$, the Bethe roots of the ground state distribute continuously.
Adding or removing a boson corresponds to adding or removing a Bethe root at the position with the largest real part on the Bethe root distribution curve.
Therefore, we have
\begin{equation}
\mu_+=\Re(\beta_{N+1}), ~~~~ \mu_-=\Re(\beta_{N}).
\end{equation}
When $t>t_o$, we have $\Re(\beta_{N+1})=\Re(\beta_{N})$ and, thus, $\Delta \mu=0$. When $t <t_o$, $\Delta \mu=\Re(\beta_{N+1})-\Re(\beta_{N})>0$.
Since $t_o$ ($U/t_o$) defines the position where the gap opens [see Fig.\ref{fig1} (d)], the Mott-Superfluid phase transition point can be determined from $U/t_o$ in the thermodynamic limit.

In addition to the Bethe root patterns, the quantum phase transition is also manifested in the ground state energy $E_g$ as a function of $U/t$ (or equally $t/U$).
We demonstrate the change of $E_g$ and the second-order derivative of $E_g/N$ versus the scaled interaction $t/U$ in Fig.\ref{fig1} (e),
where $E_g$ shows a steep descent when $t$ exceeds a specific value $t_m$, accompanied by a peak in $-(d^2 E_g/N d t^2)$ at $t_m$.
With the finite-size scaling of the maximum of ($d^2 E_g/N d t^2$) shown in  Fig.\ref{fig1} (f),
we get $\left.(d^2 E_g/N d t^2)\right|_{t=t_m}\propto \sqrt{N},$
indicating that the second-order derivative of $E_g/N$ at $t=t_m$ is divergent in the thermodynamic limit.

The finite-size scaling behavior of $t_o$ and $t_m$ is depicted in Fig.\ref{fig1}(g),
and it is observed that the critical points determined by the analysis of BAEs and the divergence point of the ground state energy are nearly identical
(with $1/N\to 0,\ |t_m-t_o|<10^{-5}$).
From this, we can extrapolate that the value of the superfluid-Mott phase transition point is $t_c= 0.17155 \pm 10^{-5}$ ($\alpha_c = U/t_c \approx 5.83$).
In addition, numerical calculations of the ground state correlation functions provide further support for this result (see Supplemental Material \cite{SM}).
It is noteworthy that this value is distinct from that in the Bose-Hubbard model with filling $\rho=1$ where
the superfluid-insulator phase transition point is $U_c=3.28\pm 0.04$ obtained by the quantum Monte Carlo simulation \cite{54}.
For the case with filling $\rho=2$, similar analysis on the distribution of Bethe roots of the ground state and the ground state energy can be carried out \cite{SM}.  Our result unveils that the value of superfluid-Mott phase transition point is $t_c= 0.10100\pm10^{-5}$.

\begin{figure}[tbp]
    \includegraphics[width=0.49\textwidth]{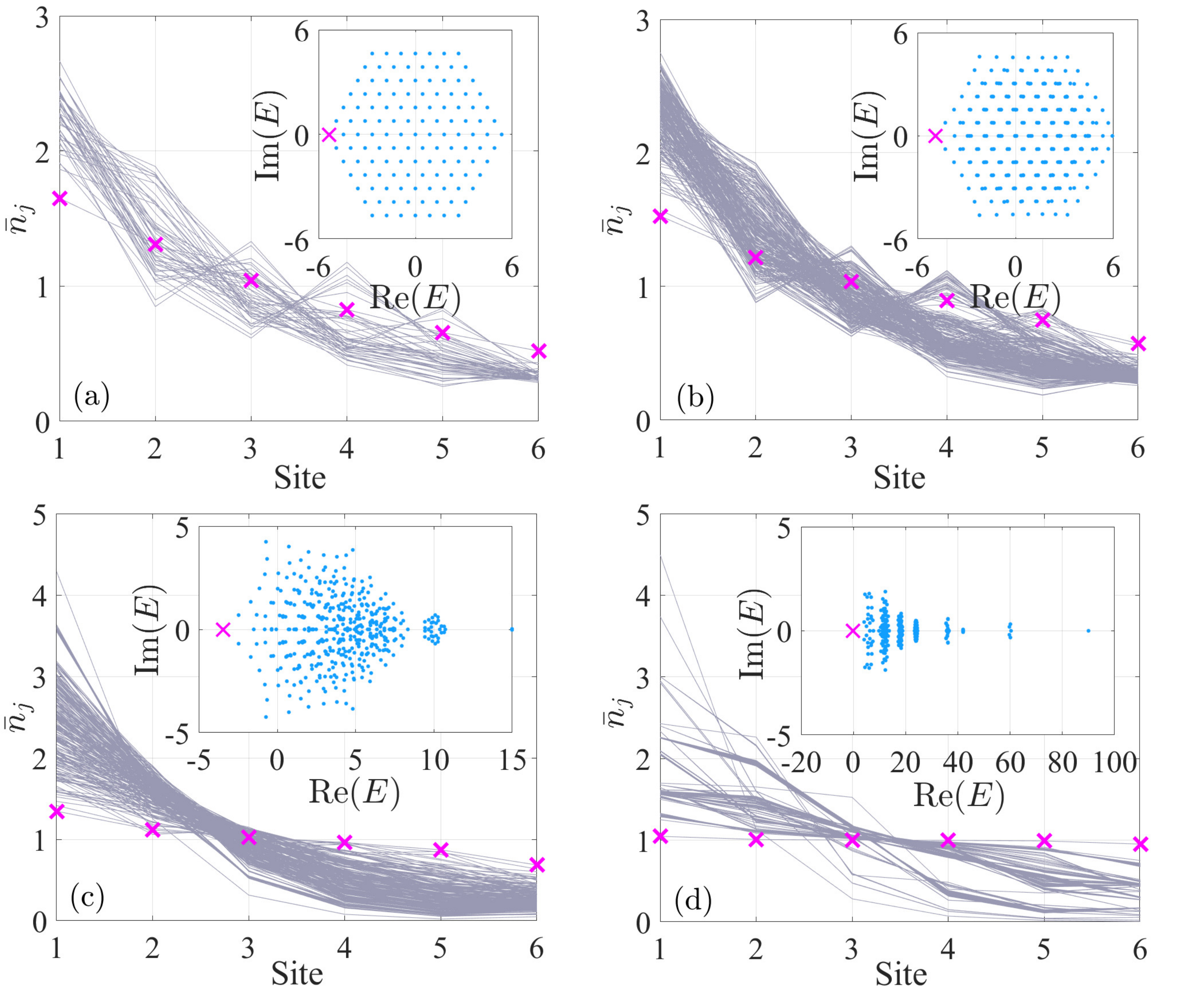}
    \caption{Exact diagonalization results of the expected values of particle number at each site $\bar{n}_j$ of all eigenstates and the corresponding eigenvalues with $U=0$ (a), 0.2 (b), 1 (c), 6 (d) respectively.
    Each point on the complex plane represents an eigenvalue, while each line corresponds to the particle number distribution of a single state. The ground state particle number distribution and the corresponding energies are marked with crosses. Common parameters: $N=6$, $M=6$, $t=1$, and $\epsilon=0.5$.}
    \label{fig2}
\end{figure}
{\it Non-Hermitian skin effects.---}
Now, we investigate the effect of boundary parameter $\epsilon$ on the eigenstates of the system (\ref{H1}) by considering $\epsilon \neq 1$, which breaks the translation invariance and leads to scale-free NHSE.
To comprehend the boundary-induced NHSE in the lattice with unidirectional hopping, it is instructive to see the noninteracting limit of $U=0$ \cite{Guo-CPB2023}, in which
all eigenstates accumulated asymmetrically near the boundaries and exhibit NHSE [see Fig.\ref{fig2} (a)].
From the BAEs~\eqref{BAE2}, for the Bethe roots pattern or the energy spectrum, the parameter value $\{t, \epsilon\}$ is equivalent to $\{t/\sqrt[N]{\epsilon}, 1\}$.
For a finite nonzero $\epsilon$, when $N\to \infty$,  $\sqrt[N]{\epsilon} \to 1$ and all the eigenvalues are equal to those of the periodic boundary case.
It means that $\epsilon$ does not change the critical point $t_c$ in the thermodynamic limit.

Next we demonstrate that the NHSE exists in the interacting system based on the exact solution.
Using the coordinate Bethe ansatz method, the eigenstates of the system can be written out explicitly
\begin{eqnarray}\label{Psi}
\ket{\Psi}_m=\sum_{x_1,..., x_M=1}^N \psi (x_1,...,x_M) b_{x_1}^{\dagger} ... b_{x_M}^{\dagger}\ket{0},
\end{eqnarray}
where the subscript $m$ takes the values from 1 to the dimensional of Hilbert space $(N+M-1)!/$$[(N-1)!M!]$, $\psi(x_1,..., x_M)$ is the wave function
\begin{eqnarray}\label{psi}
\psi =\sum_{p,q} A_p (q) \prod_{j=1}^M \Big(\frac{\beta_{p_j} }{-t}\Big)^{x_{q_j}}\theta (x_{q_1}\leq ... \leq x_{q_M})
\end{eqnarray}
with $p=\{p_1,...,p_M\}$ and $q=\{q_1,...,q_M\}$ being the permutations of $\{1,..., M\}$, $\theta (x_{q_1}\leq ... \leq x_{q_M})$ is the generalized step function,
which is one in the noted regime and zero in the others, and $\ket{0}$ is the vacuum state. The amplitude $A_p$ satisfies
\begin{equation}\label{Ap}
\frac{A_{p_1... p_{j+1} p_j ... p_M}}{A_{p_1... p_j p_{j+1} ... p_M}}=\frac{\beta_{p_{j+1}}-\beta_{p_{j}}+U}{\beta_{p_{j+1}}-\beta_{p_{j}}-U}.
\end{equation}
Each set of Bethe roots $\{\beta_j\}$ of BAEs~\eqref{BAE2} completely determines one eigenstate.

The expected values of particle number operator at the $j$th site $\bar{n}_j={}_m \bra{\Psi} n_j \ket{\Psi}_m$
for all the eigenstates are shown in Fig.\ref{fig2} (b)-(d), at different values of $U$ while $t$ is fixed to 1.
It is shown that the distributions for most eigenstates are localized asymmetrically around the boundary, exhibiting the characters of NHSE.
This can be understood directly by analyzing the properties of wave functions.
Multiplying the $M$ BAEs \eqref{BAE2}, we have
\begin{equation}\label{betaepsilon}
\prod_{j=1}^M |\frac{\beta_j}{t}|=\epsilon^{M/N},
\end{equation}
which gives that
\begin{eqnarray}
 |\psi (x_1+1,..., x_M+1)|=\epsilon^{M/N}|\psi (x_1,...,x_M)|,
\end{eqnarray}
where $x_j<N$. For the periodic system with $\epsilon=1$, the wave function has translational symmetry so that the particle number distribution is uniform.
With $\epsilon \neq 1$, the translational symmetry is broken.
Because of the factor $\epsilon^{M/N}$, the wave function will decay if $\epsilon<1$ or increase if $\epsilon>1$ along the sites.  Localization lengths of the skin states at $U=0$ are proportional to the system size and are actually the so-called scale-free localized states \cite{LiLH2021}.
The property~\eqref{betaepsilon} is independent on the value of $t/U$.
As shown in  Fig.\ref{fig2} (b) and (c), the NHSE still survives when interaction is considered.
When $U$ is large enough, the Mott phase emerges and the skin effect of the ground state is suppressed as shown in  Fig.\ref{fig2} (d).
Particularly, the density distribution of a Mott state is uniformly distributed in the thermodynamical limit, and, thus, the NHSE is completely suppressed.
This can be manifested from our numerical results (see Supplemental Material \cite{SM}) and also the analytical expression of the ground state wave function in the limit of $U \rightarrow \infty$ given by
$
    \ket{\Psi}_G=b_{1}^{\dagger} b_{2}^{\dagger} ... b_{N}^{\dagger}\ket{0}.
$
In \cite{SM}, we numerically calculate the average deviation of the uniform distribution defined as $\delta=\frac{1}{N} \sum_{j=1}^N |\bar{n}_j-\rho|$. Finite-size analysis of our numerical results indicates that $\delta$ approaches zero in the limit of $N \rightarrow \infty$ for  $t/U \leq 0.17$, whereas it approaches a nonzero value for  $t/U\geq 0.18$. These numerical results suggest that the skin effect of a Mott state is completely suppressed in the thermodynamical limit.

{\it Summary and outlook.--}
In the framework of the quantum inverse scattering method, we proved the integrability of the Bose-Hubbard model with unidirectional hopping and derived the Bethe ansatz equations for the system.
Based on the exact solution, we investigated the energy spectrum and distribution of Bethe roots of the ground state, which allows us to identify the existence of a superfluid-Mott insulating transition for integer fillings. Furthermore, we studied the effect of boundary parameter $\epsilon$ on the system and demonstrated the presence of NHSE when $\epsilon$ departs from $1$. We also elucidated the effect of interaction on NHSE and showed that the NHSE is completely suppressed for the Mott insulator state in the thermodynamical limit.
Our work introduces a new class of integrable models without Hermitian correspondence and sheds light on the interplay between interaction and NHSE. Additionally, our study provides a benchmark for testing numerical techniques developed for non-Hermitian many-body systems.

Theoretically, it has been unveiled that the combination of loss and multichannel interference with tunable phases can generate nonreciprocal hopping and even unidirectional transmission \cite{LiuYC}.
Dissipative chain model with nonreciprocal hopping has been implemented in cold atomic optical lattices \cite{ExpColdLiang}, where nonreciprocal hopping is effectively generated by introducing an auxiliary lattice with on-site loss \cite{ExpColdLapp,ExpColdLiang,LiuYC}. Therefore, the Bose-Hubbard model with unidirectional hopping is principally realizable in current cold atomic experiments.

\begin{acknowledgments}
We thank Yanxia Liu for helpful discussions on the numerical study of non-Hermitian models and Kang Wang for his support in density matrix renormalization group numerical calculations.
The financial supports from National Key R$\&$D Program of China (Grant No. 2021YFA1402104),
the National Natural Science Foundation of China (Grants No. 12074410, No. 12174436, No. 11934015,
and No. T2121001), and Strategic Priority Research Program of the Chinese Academy of Sciences (Grant No. XDB33000000) are gratefully acknowledged.

\end{acknowledgments}

\onecolumngrid
\newpage
\renewcommand{\theequation}{S\arabic{equation}}
\renewcommand{\thefigure}{S\arabic{figure}}
\renewcommand{\thetable}{S\arabic{table}}
\setcounter{equation}{0}
\setcounter{figure}{0}
\setcounter{table}{0}

\begin{center}
    {\bf \large Supplemental Material for ``Exact solution of  Bose-Hubbard model with unidirectional hopping" }
\end{center}

\section{Derivation of Bethe Ansatz equations by the algebraic Bethe ansatz method}

In this section of the supplementary material, we present the detailed procedures of deriving the Bethe ansatz equations (BAEs) by the algebraic Bethe ansatz method \cite{Korepin,WYCS}.

We first consider the $R$ matrix $R_{0,\bar{0}}(u) \in \text{End}(V_0 \otimes V_{\bar{0}})$ given by
\begin{equation}
	R_{0,\bar{0}} (u)=\begin{pmatrix}
	u-\eta& 0 & 0 & 0\\
	0 & u & \eta & 0\\
	0 & \eta & u & 0\\
	0 & 0 & 0 & u-\eta \end{pmatrix},
	\end{equation}
where $u$ is the spectral parameter, $\eta=-1$, $V_0 $ and $ V_{\bar{0}}$ represent $2\times2$ auxiliary spaces.
Then we introduce the Lax operator \cite{Kuznetsov89}. In the auxiliary space, the Lax operator is written as
\begin{equation}
	L_{0,j}(u)=\begin{pmatrix}
	u-n_{j} &g b_{j}\\g b_{j}^{\dag}& -g^2
	\end{pmatrix}.
	\end{equation}
It can be verified that $R$ matrix and Lax operator satisfy the Yang-Baxter equation
\begin{equation}\label{a-1}
	R_{0,\bar{0}} (u-v) L_{0,j} (u) L_{\bar{0},j} (v)=L_{\bar{0},j} (v) L_{0,j} (u) R_{0,\bar{0}} (u-v).
\end{equation}

The monodromy matrix $T(u)$ is constructed as the product of the $L$ operators at $N$ different sites of one chain and a $c$-number matrix $K_0$:
\begin{equation}
	\begin{split}
	T_0(u)=L_{0,N}(u) L_{0,N-1}(u) \cdots L_{0,1}(u) K_0.\\
	\end{split}
\end{equation}
In the auxiliary space, $K_0$ is a diagonal matrix given by $ K_0= diag(1, \epsilon)$.

From the Yang-Baxter equation (\ref{a-1}), we can further derive the matrix $T(u)$ satisfies the Yang-Baxter relation (the $RTT$ relation)
\begin{equation}\label{a-2}
	R_{0,\bar{0}} (u-v) T_{0} (u) T_{\bar{0}} (v)=T_{\bar{0}} (v) T_{0} (u) R_{0,\bar{0}} (u-v).
\end{equation}
Using the $RTT$ relation \eqref{a-2}, it can be proven that the transfer matrices with different spectral parameters commute with each other, i.e., $[ t(u), t(v)]=0$.
Taking the trace over the auxiliary space of monodromy matrix, we obtain the transfer matrix
\begin{equation}
	\begin{split}
	t(u)&=\text{tr}_0[L_{0,N}(u) L_{0,N-1}(u) \cdots L_{0,1}(u) K_0]\\
	&= u^{N}+C_{1}  u^{N-1}+C_{2} u^{N-2}+\cdots.
\end{split}
\end{equation}
The transfer matrix $t(u)$ is an \textit{N}th-degree polynomial with respect to $u$ which gives $N$ different conserved quantities $C_n$ ($n = 1, ..., N$).
Direct calculation gives the operators $C_1$ and $C_2$ as
\begin{eqnarray}
&&C_{1}=-\sum_{j=1}^{N} n_{j}, \\
&&C_{2}= \frac {1}{2} \sum_{i\neq j}^{N} n_{i} n_{j}+g^2 \left(\sum_{j=1}^{N-1} b_{j}^{\dag}b_{j+1} + \epsilon b_{N}^{\dag}b_{1}\right).
\end{eqnarray}
The integrable Hamiltonian is constructed as
\begin{equation}
H=\frac{t}{2g^2}\left(C_1^2+C_1-2 C_2 \right),
\end{equation}
where the on-site interaction is parameterized as $U=t/g^2$. It is easy to see $[H, C_n] = 0$, so that $H$ is integrable.

Next we need to diagonalize the transfer matrix to get the eigenstates and eigenvalues with the help of the $RTT$ relation (\ref{a-2}).
In the auxiliary tensor space, the monodromy matrix can be expressed as
\begin{equation}
	T(u)=\begin{pmatrix}
		A(u) & B(u)\\
		C(u) & D(u) \end{pmatrix},
\end{equation}
where $A (u)$, $B (u)$, $C (u)$, $D (u)$ are operators depending on the spectral parameter $u$. Then the transfer matrix is
\begin{equation}
	t(u)=A(u)+D(u).
\end{equation}
From the $RTT$ relation (\ref{a-2}),
we can get the commutation relations between these operators:
\begin{equation}
    \begin{split}\label{a-3}
   [A(u),A(v)]&=[B(u),B(v)]=[C(u),C(v)]=[D(u),D(v)]=0, \\
   A(u) C(v) = &  \frac{u-v-1}{u-v}   C(v)A(u)+ \frac{1}{u-v} C(u)A(v),\\
   D(u) C(v)=& \frac{u-v+1}{u-v} C(v)D(u)-\frac{1}{u-v} C(u)D(v).
\end{split}
\end{equation}
The reference state we choose is the vacuum state
\begin{equation}
		\ket{0}=\ket{0}_1 \otimes \ket{0}_2 \otimes \dots \otimes \ket{0}_N.
\end{equation}
Applying the monodromy matrix $T(u)$ to the reference state, we have
\begin{equation}
	\begin{split}
		&A(u)\ket{0}=u^N\ket{0},\\
		&D(u)\ket{0}=\epsilon (-\frac{t}{U})^N \ket{0},\\
		&B(u)\ket{0}=0.
	\end{split}	
\end{equation}
The operator $C(u)$ contains bosonic creation operators and can be used to construct
the Bethe states
\begin{equation}
	\ket{\mu_1,...,\mu_M}=\prod_{j=1}^{M} C(\mu_j)\ket{0},
\end{equation}
where $M$ is the particle number and $\{\mu_j\}$ is a set of parameters.
Based on the relation~\eqref{a-3}, the following useful formulae can be derived:
\begin{equation}
	\begin{split}\label{a-4}
		A(u)C(\mu_1)...C(\mu_M)=&\prod_{j=1}^{M} \frac{u-\mu_j-1}{u-\mu_j} C(\mu_1)...C(\mu_M)  A(u) \\
	&+ \sum_{j=1}^{M}\frac{1}{u-\mu_j} \prod_{l\neq j}^M \frac{\mu_j-\mu_l-1}{\mu_j-\mu_l} C(\mu_1)...C(\mu_{j-1})C(u) C(\mu_{j+1})...C(\mu_M) A(\mu_j),\\
	D(u)C(\mu_1)...C(\mu_M)=&\prod_{j=1}^{M} \frac{u-\mu_j+1}{u-\mu_j} C(\mu_1)...C(\mu_M)  D(u) \\
	&- \sum_{j=1}^{M}\frac{1}{u-\mu_j} \prod_{l\neq j}^M \frac{\mu_j-\mu_l+1}{\mu_j-\mu_l} C(\mu_1)...C(\mu_{j-1})C(u) C(\mu_{j+1})...C(\mu_M) D(\mu_j).
	\end{split}	
\end{equation}

Applying $t(u)$ to the Bethe state and using relations (\ref{a-3}) (\ref{a-4}), we have
\begin{equation}
	\begin{split}\label{a-5}
t(u)\ket{\mu_1,...,\mu_M}&=\left(A\left(u\right)+D\left(u\right)\right)\prod_{j=1}^{M} C(\mu_j)\ket{0}\\
&=\Lambda (u) \prod_{j=1}^{M} C(\mu_j)\ket{0}+ \sum_{j=1}^{M} \Lambda_j (u) C(\mu_1)...C(\mu_{j-1})C(u) C(\mu_{j+1})...C(\mu_M)\ket{0},
	\end{split}
\end{equation}
where
\begin{equation}
	\begin{split}\label{a-6}
		&\Lambda(u)=u^N \prod_{j=1}^{M}\frac{u-\mu_{j}-1}{u-\mu_{j}}+\epsilon (-\frac{t}{U})^N \prod_{j=1}^{M}\frac{u-\mu_{j}+1}{u-\mu_{j}},	\\
		&\Lambda_j (u)=\frac{1}{u-\mu_j}\left(\mu_j^N \prod_{l\neq j}^M \frac{\mu_j-\mu_l-1}{\mu_j-\mu_l}-\epsilon (-\frac{t}{U})^N \prod_{l\neq j}^M \frac{\mu_j-\mu_l+1}{\mu_j-\mu_l}  \right).
	\end{split}
\end{equation}
To ensure the Bethe state to be an eigenstate of the transfer matrix, the unwanted terms must vanish, i.e., $\Lambda_j (u) = 0$. This induces the BAEs:
\begin{equation}
	\frac{U^N}{(-t)^N}\mu_j^N=\epsilon  \prod_{l\neq j}^M \frac{\mu_j-\mu_l+1}{\mu_j-\mu_l-1}, \quad j=1,\cdots, M.
\end{equation}
From \eqref{a-4} and \eqref{a-5}, we can see that the solutions \{$\mu_j| j = 1, . . . , M$\} should satisfy $\mu_j \neq \mu_l$ for $j\neq l$.

Expanding $\Lambda (u)$ and comparing with the polynomial of $t(u)$, we can get
\begin{equation}
	\begin{split}
		&C_1 \ket{\mu_1,...,\mu_M}= -M\ket{\mu_1,...,\mu_M},\\
		&C_2\ket{\mu_1,...,\mu_M}=\left(-\sum_{j=1}^M \mu_j+\frac{M(M-1)}{2}\right)\ket{\mu_1,...,\mu_M}.
	\end{split}
\end{equation}
Thus, the eigenvalue of Hamiltonian is
\begin{equation}
E=U \sum_{j=1}^M \mu_j.
\end{equation}

For convenience, we put $\beta_j=U \mu_j$. The BAEs can be rewritten as
\begin{equation}\label{BAE1}
		\left(\frac{\beta_j}{-t}\right)^N= \epsilon \prod_{l\neq j}^M \frac{\beta_j-\beta_l+U}{\beta_j-\beta_l-U}, \quad j=1,\cdots, M,
\end{equation}
where $\beta_j \neq \beta_l$ for $l\neq j$. The eigenvalue of the Hamiltonian becomes
\begin{equation}
E=\sum_j^M \beta_j.
\end{equation}

\section{Solutions of Bethe roots}

Here we provide more detailed analysis of BAEs~\eqref{BAE1}.
We consider the case of $\epsilon=1$ and mainly concern the Bethe root patterns in the two limits $t/U\to \infty$ and $t/U\to 0$.


When $U$ is small ($t/U$ is large), the right side of BAEs~\eqref{BAE1} tends to 1, so that
the solution $\{\beta_j\}$ is
\begin{equation}
\beta_j=te^{\frac{i\pi(N+2 l)}{N}}+\delta_{j}, l=1,\cdots, N, j=1,\cdots, M,\label{Uapprox10}
\end{equation}
where $\delta_{j}$ is a perturbation induced by $U$. When $U=0$, $\delta_{j}=0$ and $\beta_j$ is the solution for the free boson with unidirectional hopping.
The corresponding spectrum, which is the sum of each set of Bethe roots, resembles regular polygons.

As $U$ becomes large ($t/U$ approaches zero), the interaction term begins to dominate and causes a change in the Bethe roots pattern.
If there exists a Bethe roots $\beta_j \in \{\beta_j\}$ satisfying $|\beta_j|>t$, the left side of BAEs \eqref{BAE1} tends to infinity in the thermodynamic limit, i.e., $|\beta_j/t|^N\to \infty$,
which requires that the denominator of right hand side should tend to zero.
Therefore, there must exist another Bethe root $\beta_k \in \{\beta_j\}$ with $\beta_k=\beta_j-U$.
If $|\beta_k|<t$,  both sides of BAEs \eqref{BAE1} tend to zero with $N\to \infty$,
and the BAEs can be satisfied.  
Conversely, if $|\beta_k|>t$, there must exist another Bethe root $\beta_l \in \{\beta_j\}$ with $\beta_l=\beta_k-U$.
Based on singularity analysis, we know that for a given $M$, there must at least exist one $\beta_{j,1}$, which satisfies $|\beta_{j,1}|<t$ in each set of solutions.
Then we conclude that if $U$ is large enough, the Bethe roots form the strings along the real axis with a length of $k$,
\begin{equation}\label{NM12}
	\beta_{j,k}=\beta_{j,1}+(k-1)U, \quad k=1,\cdots, M,
\end{equation}
here $|\beta_{j,k}/U|\approx k-1$.
Since the quantum number $I_j$ corresponding to each Bethe root has only $N$ different values, $\beta_{j,1}$ has $N$ different values, i.e. there are $N$ different strings.

The string structure reflects the unique properties of bosonic systems. When the interaction strength $U$ is large, the hard-core boson exclusion principle leads to the restriction of $\beta_{j,1}$ to $N$ distinct values. The Bethe roots form strings with gaps $U$, which describe the potential energy required to add a boson to one site already occupied by $k$ bosons.

In this way, the pattern of Bethe roots can be determined. For a given $M$, suppose that there are $m_l$ Bethe roots in the $l$-th string
\begin{equation}
	\sum_{l=1}^N m_l=M,\ l=1,...,N.
\end{equation}
Each set of $\{m_l\}$ corresponds exactly to one solution of BAEs \eqref{BAE1}
\begin{equation}
		\{\beta_j\}=\{\beta_{1,1},\beta_{1,2},...,\beta_{1,m_1},\beta_{2,1},...,\beta_{2,m_2},...,\beta_{N,1},...,\beta_{N,m_N}\}.
\end{equation}
The corresponding energy is $E\approx\sum_{l=1}^{N} m_l(m_l-1)U/2$.

\begin{figure*}[ht]
	\includegraphics[width=0.60\textwidth]{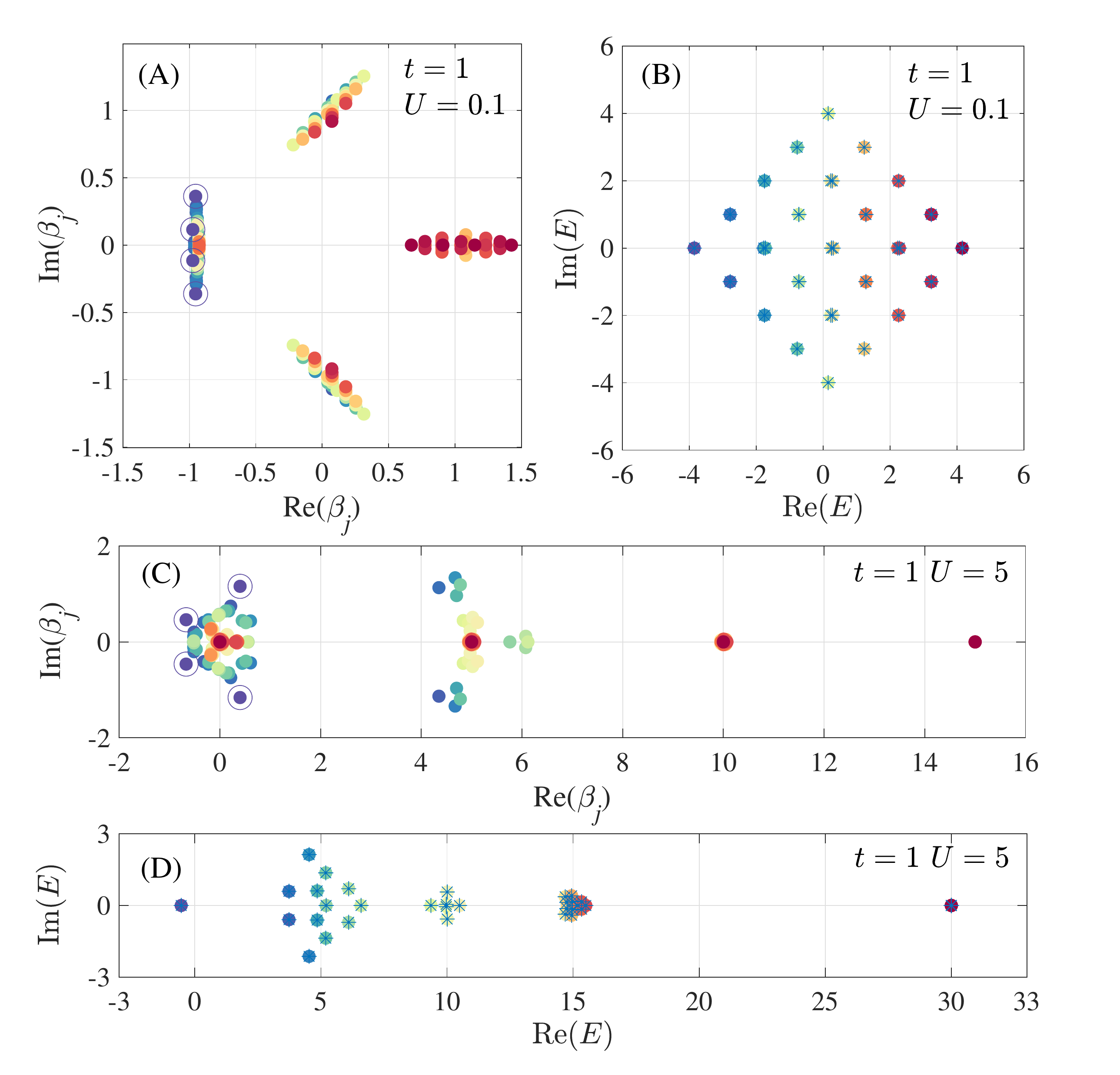}
	\caption{The distributions of Bethe roots and the energy spectrum with $U=0.1$ in (A, B), $U=5$ in (C, D).
		Each set of Bethe roots is color-coded in (A, C), with corresponding eigenvalues marked in (B, D) using the same color. The hollow circles denote the Bethe roots for the ground state, while blue asterisks represent exact diagonalization results. Common parameters: $N=4$, $M=4$, $t=1$, $\epsilon=1$.  }
	\label{figS1}
\end{figure*}

The numerical verification of above results are given in Fig.\ref{figS1}. By solving the BAEs \eqref{BAE1} under different values of $U$, the corresponding Bethe roots are obtained (Figs.\ref{figS1}(A) and (C)).
When $t/U$ is large, all $\beta_j$s are distributed around $t\exp[i\pi(N+2 l)/N]$ ($l=1,\cdots,N$). When $t/U$ is small, all $\beta_j$s are localized around discrete regimes satisfying $\Re(\beta_j)/U\approx k-1$ $(k=1,...M)$.
Based on the above Bethe roots, we obtain the spectrum of the system and show them in Fig.\ref{figS1}(B) and (D).
The spectrum can also be calculated by using the numerical exact diagonalization method (shown in Fig.\ref{figS1}(B) and (D) as the asterisks).
We can see that the results obtained by the two methods are in good agreement with each other.

We define the ground state energy as the one with minimum real part.
It is worth noting that the $M$ Bethe roots at the ground state are always symmetrically distributed around the real axis, ensuring that the ground state energy is always real.
When $t/U\to \infty$, the excitation of the system is gapless, which is independent of the filling factor $\rho=M/N$.
When $t/U \to 0$, the solutions of $\beta_j$s are described by the string solution.
The solution corresponding to the ground state is the one among all possible solutions with the smallest sum of the real parts of the roots.
Thus, the solution is $\{\beta_{j,k}\}$ with $k$ as small as possible.
Considering that each $k$ layer in the string structure has at most $N$ possible positions, the pattern of the solution of the ground state is related to $\rho$.
Especially, when $\rho=1$, the solution $\{\beta_j\}$ correspondings to the ground state is composed of $M$ different $\beta_{j,1}$.
In this case, the right side of BAEs~\eqref{BAE1} tends to $(-1)^{M-1}$, so that
\begin{equation}\label{NM13}
	\beta_{j} =te^{\frac{i\pi (N+M+2l-1)}{N}}, l=1,\cdots, N, j=1,\cdots, M.
\end{equation}
Now, adding one boson to this system results in a change of the ground state energy by $\mu_+=\Re{(\beta_{N+1})}\approx U$, while removing one boson results in a change of the energy by $\mu_-=\Re{(\beta_N)}=t$. We have $\mu_+>\mu_-$.
It is evident that this phenomenon of energy gap arising from particle addition or removal emerges only in systems with integer fillings.

\section{More numerical analysis on the Superfluid-Mott Phase Transition}

In this section, we present numerical analysis on the superfluid-Mott phase transition based on the calculation of density-density correlation functions.
Using the non-Hermitian density matrix renormalization group (DMRG) algorithm, we calculate the correlation function $C(r)$  of the ground state, which is defined as
\begin{equation}
	C(r)=\braket {n_{r}n_0}-\braket{n_{r}}\braket{n_{0}}.
\end{equation}
In references \cite{Haldane81,Giamarchi81,Kuhner00,Ejima11}, it has been unveiled that the density-density correlation function exhibits a power-law decay with distance the superfluid state, whereas it follows an exponential decay in the Mott phase.

\begin{figure}[h]
	\includegraphics[width=0.75\textwidth]{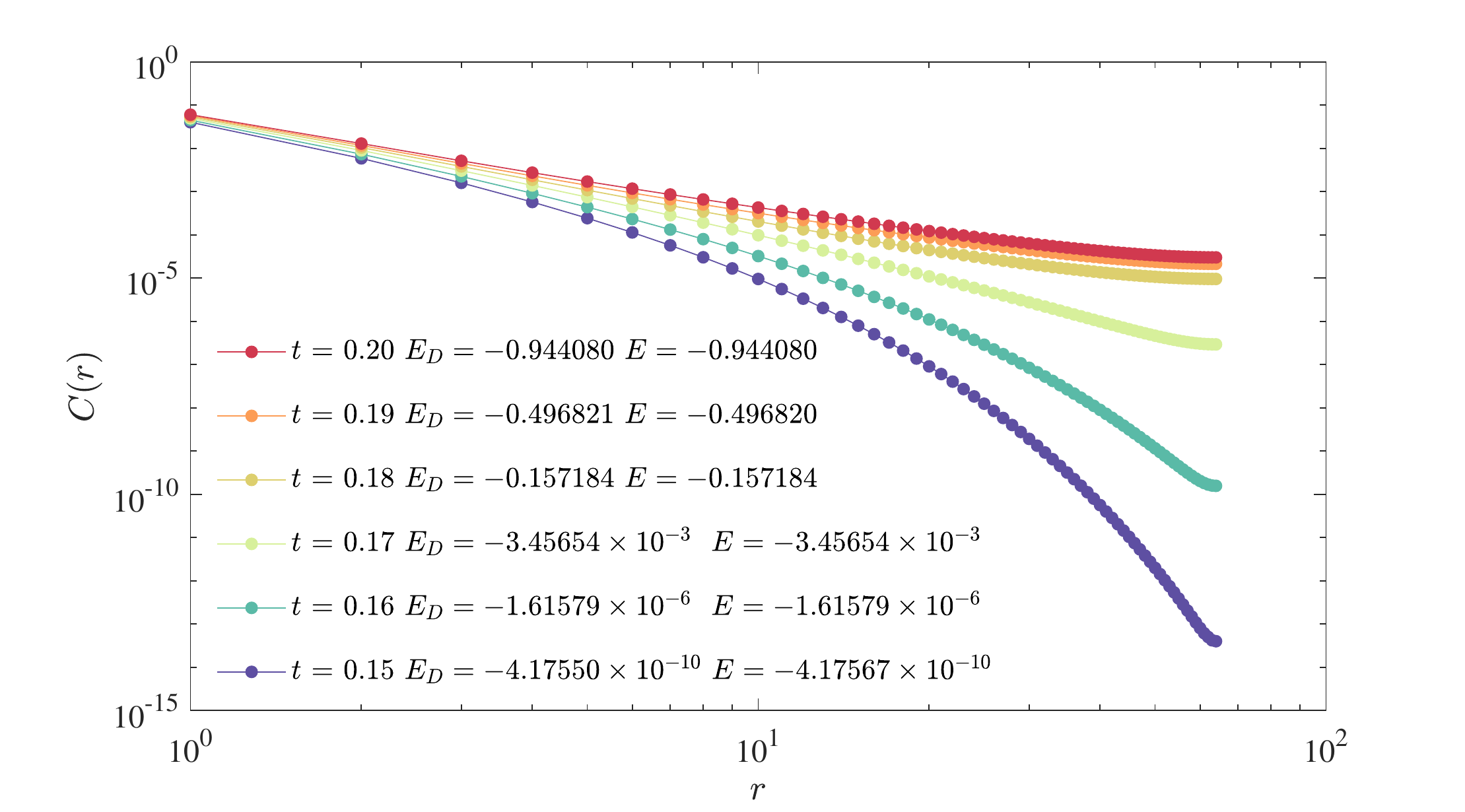}
	\caption{Density-density correlation function $C(r)$ calculated by the non-Hermitian DMRG algorithm with periodic boundary condition $\epsilon=1$.
	$E_D$ is the ground state energy calculated by the non-Hermitian DMRG algorithm and $E$ corresponds to the ground state energy obtained from the exact solution. It can be observed that the ground state energy values provided by these two methods are nearly identical.
	Here, we fix $N=M=128$, $U=1$, $\epsilon=1$.}
	\label{figS2}
\end{figure}

In Fig.\ref{figS2}, we display the correlation function $C(r)$ versus $r$ calculated by the non-Hermitian DMRG algorithm under the periodic boundary condition for various values of $t$ around the transition point.
We compare the ground state energy obtained by the non-Hermitian DMRG algorithm with the exact result obtained by Bethe ansatz and find they agree very well.
From Fig.\ref{figS2}, we can observe that at $t/U=0.18$, 0.19, and 0.20, the behavior of the $C(r)$ curves in terms of decay is entirely distinct from that observed at $t/U=0.15$ and 0.16.
The upper three curves in the graph either approach or decay more slowly than a power-law, while the lower two curves significantly deviate from the linear behavior observed in the double-logarithmic plot, indicating a faster decay rate than a power-law.
At $t/U=0.17$, the decay pattern closely aligns with and slightly exceeds a power-law decay rate.
It is noticeable that the critical point for the superfluid-to-Mott phase transition in the system is slightly greater than $t/U=0.17$. This observation is consistent with our previous conclusion.
\begin{figure}[ht]
  \includegraphics[width=1\textwidth]{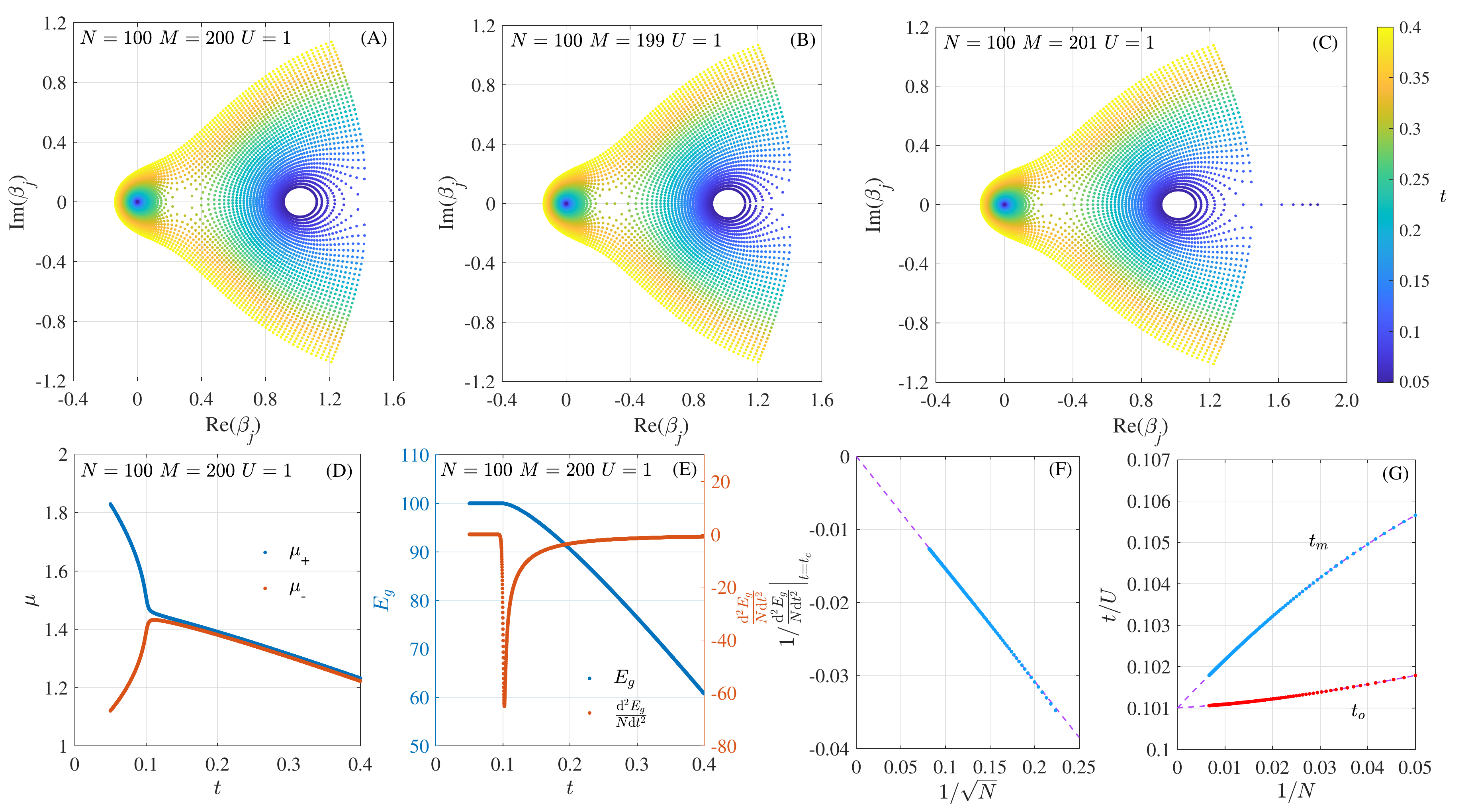}
  \caption{(A, B, C) Distribution of Bethe roots of the ground state for $N=100,\ M=\ 200\  \text{(A)},\ 199\ \text{(B)},\ 201\ \text{(C)}$
  	with $t$ varying from 0.4 to 0.05 (the step size is 0.01).
(D) Numerical results of $\mu_+$, $\mu_-$ of the system with $N=100$,$M=200$, $U=1$ with $t$ varying from 0.4 to 0.05 (The step size is 0.0001).
(E) The ground state energy (blue point) and the second derivative of $E_g/N$ with respect to $t/U$ (red point) for $N=100$, $M=200$ with $t$ varying from 0.4 to 0.05 (step size 0.0001).
(F) The extreme values of $\frac{\mathrm{d^2}E_g}{N\mathrm{d}t^2}$ plotted against $1/\sqrt{N}$ from $N=20$ to $N=150$ ($\rho=2$).
The dashed fitted lines is $1/\frac{\mathrm{d^2}E_g}{N\mathrm{d}t^2}=C_1/\sqrt{N}+C_2$. The fitting parameters are $C_1=-0.15374$, $C_2=-6.6690\times 10^{-6}$.
(G) The value of $t_m$ (blue points) and $t_o$ (red points) versus $1/N$ with $N=20$ to $150$ ($\rho=2$).
The dashed lines are fitted to determine the values of $t_m$ and $t_o$ as the system size tends to infinity.
The fitting results are $t_m=0.10100$ and $t_o=0.10101$. Here we fix $U=1$, $\epsilon=1$.
}
  \label{figS3}
\end{figure}

\section{The case of $\bf{\rho=2}$}

From the analysis of BAEs, we can see that there exists quantum phase transition in this model as long as $\rho=$ integer.
Here we discuss the case of $\rho=2$.

For the system with $M=2N$, for small values of $U$, all the Bethe roots tend to concentrate around $(-t,0)$ in the complex plane.
As $U$ increases, $N$ of these Bethe roots gradually move towards the origin $(0,0)$, while the remaining $N$ Bethe roots move towards $(U,0)$.
If a boson is removed from the system, $\beta_{2N}$ disappears, and the pattern of Bethe roots resembles that of $M=2N$.
Differently, when a boson is added to the system, the $2N$ Bethe roots in the first two layers ($k=1,2$) are already occupied, so $\beta_{2N+1}$ will be squeezed out as $U$ increases, eventually approaching $(2U,0)$.

The solutions of the BAEs \eqref{BAE1} with $M=2N$ and $M=2N\pm1$ are given in Figs.\ref{figS3} (A, B, C).
Similar to the $\rho=1$ case, there exists an odd point $t_o$ where $\beta_{2N}$ and $\beta_{2N+1}$ coincide.
The diagram of $\mu_+$ and $\mu_-$ (Fig.\ref{figS3} (D)) shows that the energy gap of the system undergoes a noticeable change near $t/U=t_o$.
Moreover, the ground state energy curve versus $t/U$ of the $\rho=2$ system has a derivative discontinuity point $t_m$ (Fig. \ref{figS3}(E)).
By analyzing the finite size scaling of the maximum of $\frac{d^2E_g}{N d t^2}$ (Fig. \ref{figS3} (F)), we observe that the second order derivative of $E_g/N$ at $t=t_m$ diverges as $N\to\infty$, with the scaling behavior given by $\left.\frac{d^2E_g}{Ndt^2}\right|_{t=t_m}\propto \sqrt{N}$. 

The finite size scaling behaviors of $t_o$ and $t_m$ are shown in Fig.\ref{figS3}(G).
The value of $t_o$ and $t_m$ converge to nearly the same value in the thermodynamic limit.
We can confirm that the superfluid-Mott phase transition critical point value is $t_c= 0.10100\pm10^{-5}$ ($\alpha_c = U/t_c \approx 9.90$) for the case of $\rho=2$.

\section{Numerical analysis on the suppression of the skin effect}
\begin{figure}[ht]
	\includegraphics[width=0.6\textwidth]{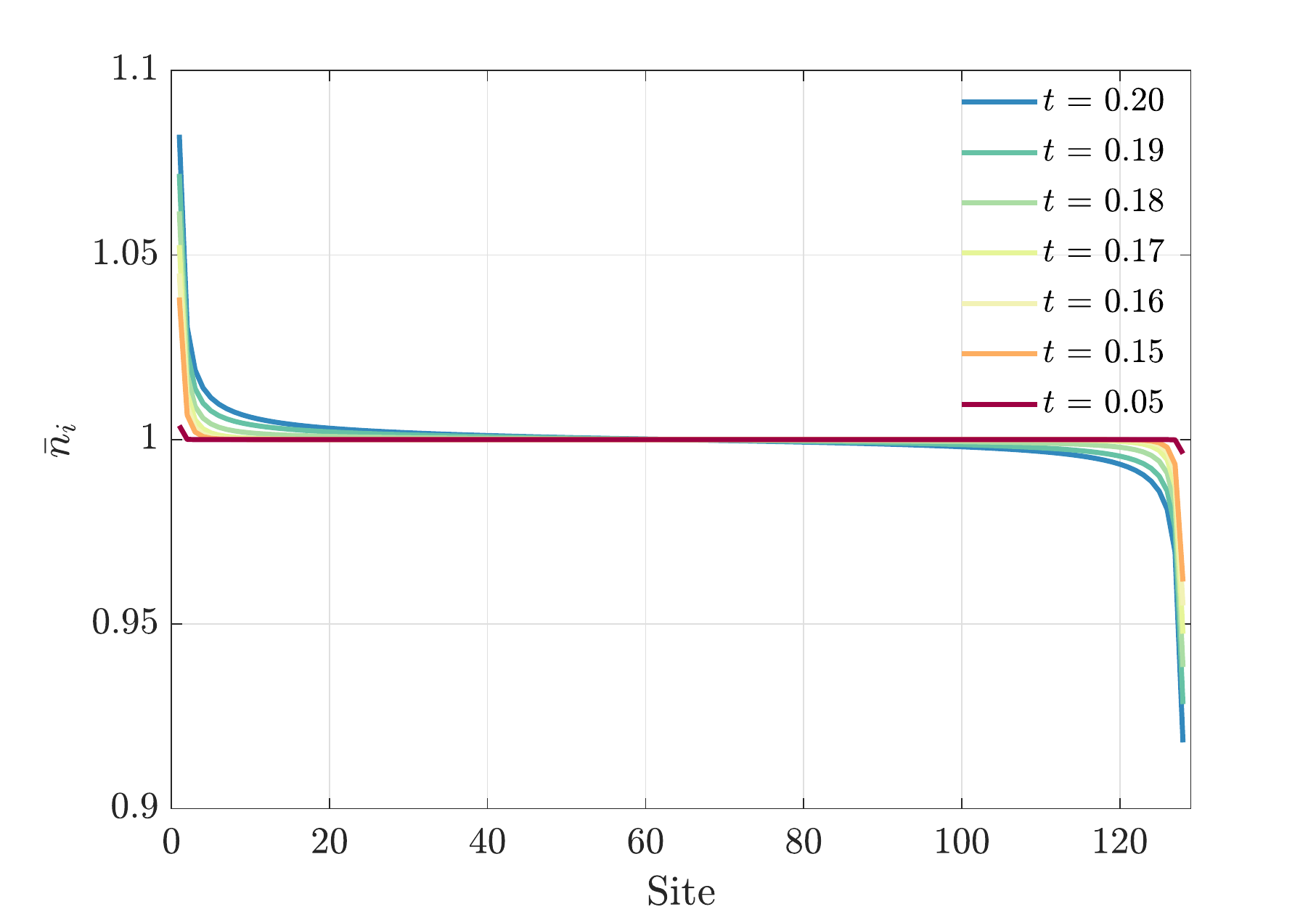}
	\caption{The density distribution of the ground state $\bar{n}_j$ with various $t$. The numerical results are obtained by using the non-Hermitian density matrix renormalization group algorithm. Here $j$ refers to the $j$-th site, and we have taken $N=128$, $\epsilon=0.5$ and $U=1$.}
	\label{figV1}
  \end{figure}

Through the analysis of the change of spectrum and the correlation, we have demonstrated that the ground state of the system undergoes a superfluid-Mott transition.
To see clearly how the density distribution changes with the change of $t/U$, we consider the case $\rho=1$ and calculate the density distribution of the ground for various $t$ by fixing $U=1$. Using the non-Hermitian DMRG algorithm, we demonstrate the density distributions of the ground state for the system with $N=128$ and various $t$ in Fig.\ref{figV1}.
Here $\bar{n}_j$ is the expected values of particle number operator at the $j$th site $\bar{n}_j= {}_{G}\bra{\Psi} n_j \ket{\Psi}_{G}$.
For fixed $N$ and $\epsilon$, the boson distribution gradually approaches a uniform distribution with decrease of the value of $t/U$.

\begin{figure}[ht]
	\includegraphics[width=0.9\textwidth]{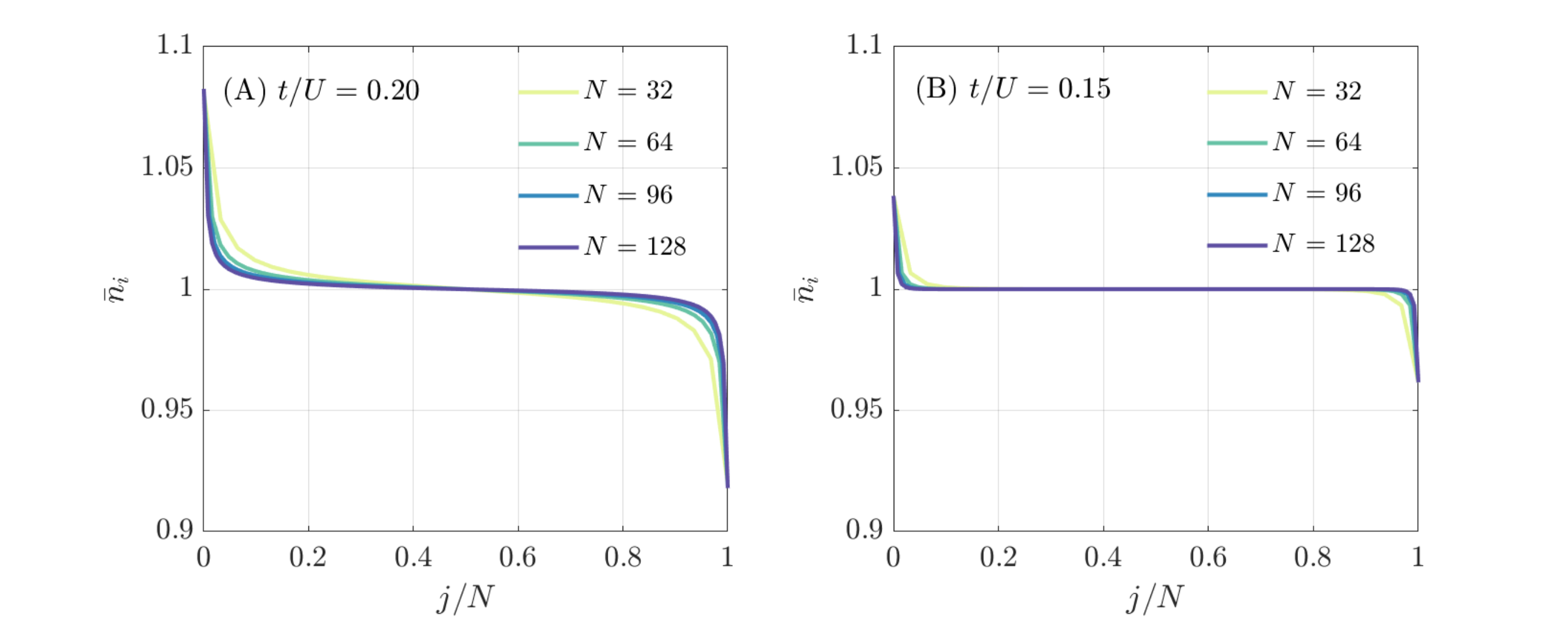}
	\caption{The density distribution of the ground state $\bar{n}_j$  with different sizes for (A) $t=0.20,U=1$ and (B) $t=0.15, U=1$. Here $j$ refers to the $j$-th site and $\epsilon=0.5$. }
	\label{figV2}
  \end{figure}

To see how the density distributions change with the lattice size $N$ in different parameter regions, we display the density distributions of systems with different sizes
in Fig.\ref{figV2}.  While Fig.\ref{figV2} (A) with $t/U=0.20$ corresponds to the superfluid phase, Fig.\ref{figV2} (B) with $t/U=0.15$ corresponds to the Mott phase.
For $t/U=0.15$, it is shown that the density distribution gradually approaches a uniform distribution with the increase of the size. However, for $t/U=0.20$, the density distribution does not approach a uniform distribution even in the large size limit.

\begin{figure}[ht]
	\includegraphics[width=0.55\textwidth]{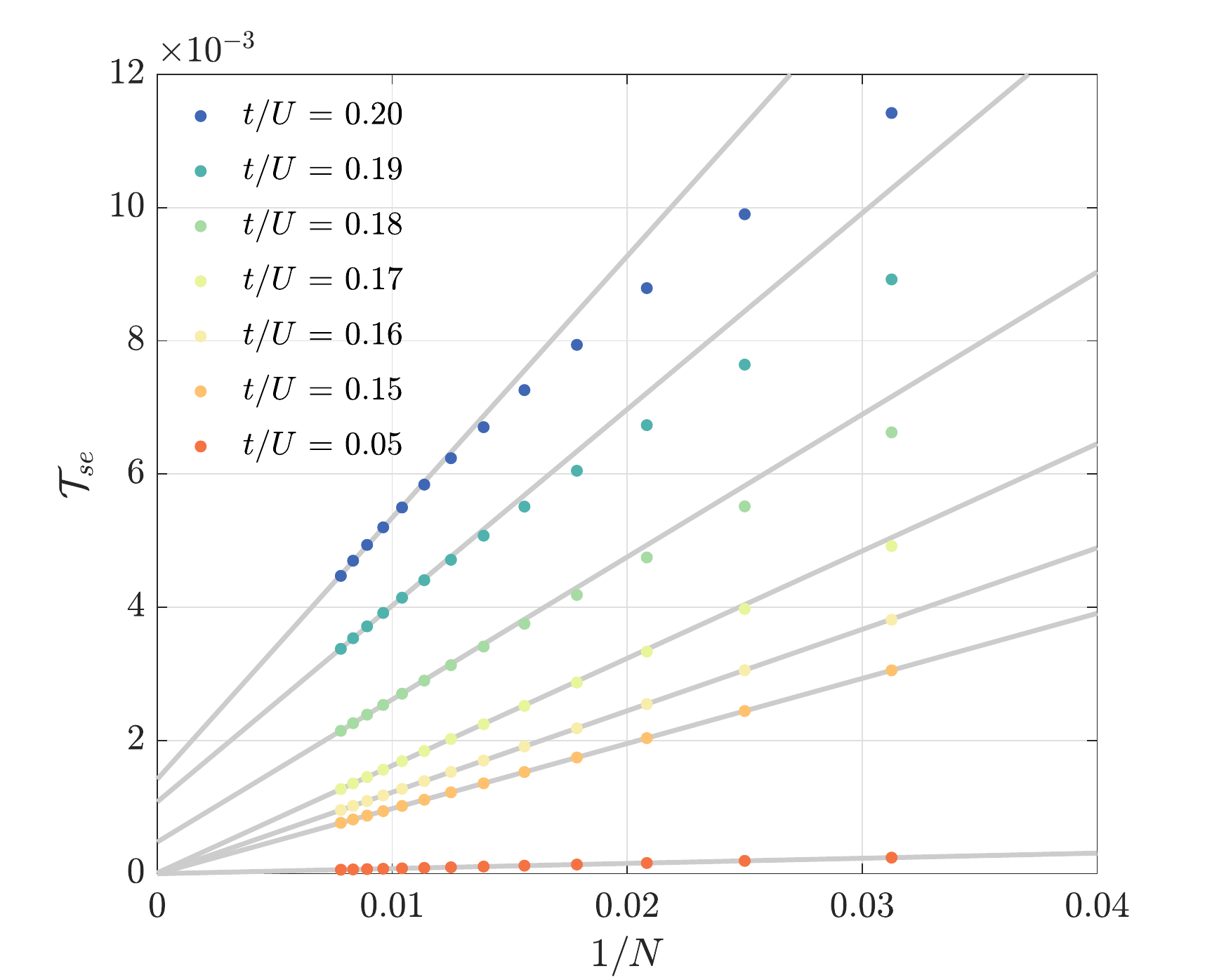}
	\caption{The value of $\delta=\frac{1}{N} \sum_{j=1}^N |\bar{n}_j-\rho|$ versus $1/N$ with $N=32$ to $128$ at intervals of $8$ under different values of $t/U$.
	The gray lines are fitted with a first-degree polynomial using data of $N=96, \ 104, \ 112, \ 120,\  128$ to determine the values of $\delta$ as the system size tends to infinity. Here $\epsilon=0.5$}
	\label{figV3}
  \end{figure}
In order to characterize quantitatively the change of density distribution around the transition point, we define
\begin{equation}
	\delta=\frac{1}{N} \sum_{j=1}^N |\bar{n}_j-\rho|
\end{equation}
to quantify an average deviation of the uniform distribution. The finite size scaling results of $\delta$ with different $t/U$ are presented in Fig.\ref{figV3}.
For $t/U \leq 0.17$, it is observed that the fitting results give $\delta = 0 $ in the thermodynamic limit.
However, for $t/U\geq 0.18$, the fitting value of $\delta>0$ is evident in the thermodynamic limit. Since $\delta =0$ implies the complete suppression of the deviation from the uniform distribution,  the skin effect of the ground state is completely suppressed in the thermodynamic limit when the system is in the Mott phase for $t/U\leq 0.17$. This is in sharp contrast with the nonzero value of $\delta$ for $t/U\geq 0.18$. The different behaviors of $\delta$ in the thermodynamic limit suggest that the critical value lies between 0.17 and 0.18, consistent with value of the superfluid-Mott phase transition given by $t_c/U = 0.17155$.

\end{document}